%% file: paper_mixed_bmst.tex
\documentclass[lettersize,journal]{IEEEtran}
\usepackage{amsmath,amsfonts}
\usepackage{amssymb}
\usepackage{algorithmic}
\usepackage{array}
\usepackage[caption=false,font=normalsize,labelfont=sf,textfont=sf]{subfig}
\usepackage{textcomp}
\usepackage{stfloats}
\usepackage{url}
\usepackage{verbatim}
\usepackage{graphicx}
\def\BibTeX{{\rm B\kern-.05em{\sc i\kern-.025em b}\kern-.08em
    T\kern-.1667em\lower.7ex\hbox{E}\kern-.125emX}}
\usepackage{balance}
\usepackage{tikz}
\usepackage{placeins}
\usepackage{standalone}
\usepackage{nicefrac}
\usepackage{orcidlink}
\usepackage{cleveref}
\usepackage[nocompress]{cite}

\crefformat{equation}{(#2#1#3)}
\crefrangeformat{equation}{(#3#1#4--#5#2#6)}
\crefmultiformat{equation}{(#2#1#3)}{ and~(#2#1#3)}{, (#2#1#3)}{, and~(#2#1#3)}

\crefname{figure}{Fig.}{Figs.} %
\Crefname{figure}{Fig.}{Figs.} %
\crefname{section}{section}{Sections}
\Crefname{section}{Section}{Sections}

\crefname{algorithm}{Algorithm\!}{Algorithms\!} %
\Crefname{algorithm}{Algorithm\!}{Algorithms\!} %

\usepackage[caption=false,font=normalsize,labelfont=sf,textfont=sf]{subfig}
\input{preamble_tikz.tex}

\begin{document}
\title{Mixed Block Markov Superposition Transmission Codes}
\author{Philipp~Mohr\,\orcidlink{0000-0003-4350-9969},
	Jasper~Brüggmann\,\orcidlink{0009-0007-7287-1488},
	Viet~Hoang~Le\,\orcidlink{0000-0002-7413-0401},
	and~Gerhard~Bauch,~\IEEEmembership{Fellow,~IEEE}
\thanks{This work is supported by Huawei Technologies Sweden AB. (Corresponding author: Philipp Mohr.)}
\thanks{The authors are with the Institute of Communications, Hamburg University of Technology, 21073 Hamburg, Germany (e-mail: philipp.mohr@tuhh.de; jasper.brueggmann@tuhh.de; viet.le@tuhh.de; bauch@tuhh.de).}
}

{}
\maketitle

\begin{abstract}
Block Markov superposition transmission (BMST) codes provide a flexible framework for constructing codes with near-capacity performance and low-complexity sliding-window decoding. However, existing BMST variants show contrasting performance limitations: recursive BMST (rBMST) codes suffer from error propagation but avoid high error floors, whereas non-recursive BMST codes exhibit the opposite behavior. Motivated by these complementary characteristics, we combine recursive and non-recursive components through parallel and serial concatenation, yielding mixed BMST (mBMST) codes. The proposed framework subsumes existing BMST variants and enables new BMST structures. Simulations show that these structures improve FER and BER performance with lower memory requirements than rBMST.
\end{abstract}
\FloatBarrier

\section{Introduction}
\IEEEPARstart{F}{orward} error-correcting (FEC) codes are a key enabler of modern communication systems, driven by increasingly stringent requirements on throughput, latency, reliability, and energy efficiency. This evolution is reflected in mobile communication standards, which have progressed from feedforward convolutional codes in 2G systems to turbo codes employing recursive components in 3G/4G, and to low-density parity-check (LDPC) codes in 5G New Radio (NR).

A promising direction for further performance improvement is spatial coupling, in which code blocks are connected through extended code constraints~\cite{felstrom1999time,kudekar2013spatially,ren2025edge}. Several classes of spatially coupled codes exhibit the threshold saturation effect, whereby the belief-propagation (BP) decoding threshold approaches the maximum-a-posteriori (MAP) threshold of the underlying block code~\cite{kudekar2013spatially}. BP decoding of spatially coupled codes can be efficiently implemented using a sliding-window algorithm~\cite{felstrom1999time}.

Block Markov superposition transmission (BMST) codes constitute an important class of spatially coupled codes~\cite{ma2015block,zhao2018recursive,zhaodoublyrecursive2020,li2024spatially}. In BMST, multiple \emph{basic} codewords are superimposed across successive transmission blocks using a \mbox{rate-1} convolutional component. BMST codes are attractive due to their flexible construction, multi-rate capability, and simple encoding. Furthermore, the regular graph structure provides high decoding locality, since each decoding step involves only a few neighboring blocks.

In this letter, we show that BMST codes employing recursive components (rBMST) suffer from error propagation but achieve low error floors, whereas non-recursive BMST (nBMST) codes exhibit the opposite behavior. To combine the strengths of both approaches, we propose a parallel concatenation of recursive and non-recursive components, leading to performance improvements of up to $0.14\dB$. The proposed alternating decoding schedule also offers potential implementation advantages. Furthermore, we generalize parallel and serial concatenation principles within the BMST framework, leading to the class of \emph{mixed BMST (mBMST) codes}. The proposed framework subsumes several existing constructions, including  nBMST~\cite{ma2015block}, rBMST~\cite{zhao2018recursive}, DrBMST~\cite{zhaodoublyrecursive2020}, and BiBMST~\cite{li2024spatially}.

\begin{figure}[t]
\centering
\includestandalone[mode=buildmissing]{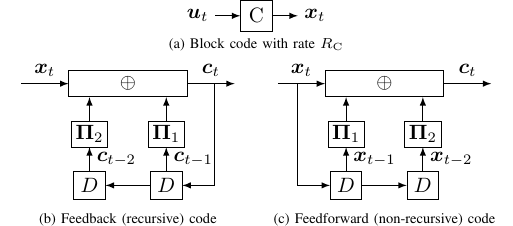}
\caption{Constituent codes for BMST (memory $m=2$ in (b) and (c)).} %
\label{fig:encoders}
\end{figure}

\section{Block Markov Superposition Transmission}
\subsection{Encoding} %
The encoding principle of BMST is depicted in \cref{fig:encoders}.
A sequence of $L$ data blocks $\uvec_t\in \mathbb{F}_2^{\KC}$ is encoded, where $t$ denotes a time index and the block length is $\KC$.
Each block $\uvec_t$ is mapped using a basic code $\CBasic[\NC,\KC]$ to a codeword $\xvec_t\in \mathbb{F}_2^{\NC}$, as depicted in \cref{fig:encoders}a.
The individual codewords $\xvec_t$ are superimposed in time using a rate-1 block-convolutional encoder with memory $m$ to form the transmitted codeword $\cvec_t\in \mathbb{F}_2^{\NC}$.
Two fundamental convolutional encoder variants exist with a feedback (recursive) structure, shown in \cref{fig:encoders}b,
\begin{align}
	\cvec_t = \coder_m(\xvec_t) \mydef \xvec_t \oplus \Pmat_1\cvec_{t-1} \oplus \ldots \oplus \Pmat_m\cvec_{t-m}
	\label{eq:rec}
\end{align}
or a feedforward (non-recursive) structure, shown in \cref{fig:encoders}c,
\begin{align}
	\cvec_t = \coden_m(\xvec_t) \mydef \xvec_t \oplus \Pmat_1\xvec_{t-1} \oplus \ldots \oplus \Pmat_m\xvec_{t-m}
	\label{eq:nrec}
\end{align}
where random permutation matrices $\Pmat_i$ are used.
The overall number of data blocks is $L$, followed by a termination sequence $\uvec_t=\myvec{0}$ for $t=L+1,\ldots,L+T$.
The rate of the BMST system is $R=\RC\cdot\frac{L}{L+T}$ with basic code rate $\RC=\nicefrac{\KC}{\NC}$.

\begin{figure}[t]
	\centering
	\includestandalone[mode=buildmissing]{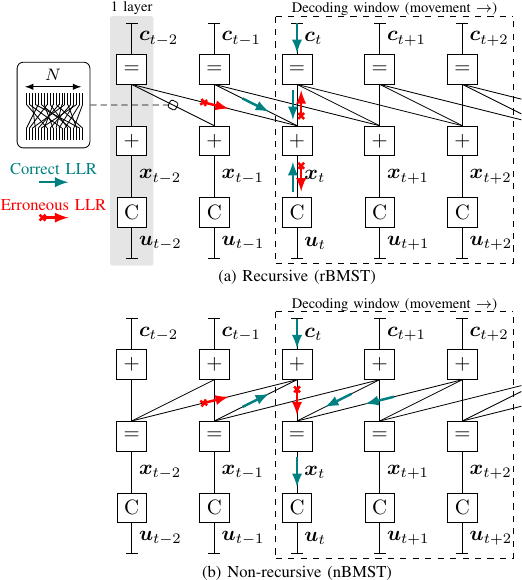}
	\caption{Error propagation in recursive and non-recursive BMST graphs.} %
	\label{fig:ng_rec_nrec}
\end{figure}

The BMST code constraint can be represented using the normal graph framework~\cite{forney2001codes}, as shown in \cref{fig:ng_rec_nrec} for the recursive encoder in \cref{eq:rec} and the non-recursive encoder in \cref{eq:nrec}, respectively.
Each edge of the graph represents a block of variables, e.g., $\cvec_t$ or $\uvec_t$.
The graph consists of repetition (REP) nodes \ngrep, single parity check (SPC) nodes \ngspc, and basic code nodes \ngbc.
The permutations $\Pmat_i$ are integrated into the edges of the graph for brevity.

\subsection{Decoding}
Each block $\cvec_t$ is modulated using binary phase-shift keying, so that $1-2\cvec_t$ is transmitted and the channel output is $\yvec_t=1-2\cvec_t+\nvec_t$.
We consider additive white Gaussian noise $\nvec_t\in\mathbb{R}^{\NC}$ with variance $\No/2$.
The decoder observes the LLRs $\lchvec_t=4\yvec_t /\No$.
A sliding window restricts processing to $W$ consecutive layers.
In \cref{fig:ng_rec_nrec}a, one layer includes the nodes \ngrep, \ngspc, and \ngbc. 
At each window position, the decoder performs $\IW$ iterations.
One iteration consists of a forward sweep and a backward sweep over all $W$ layers in the decoding window. At each visited layer, the node updates are applied sequentially according to the following schedule:
\begin{itemize}
	\item $\ngrep\to\ngspc\to\ngbc\to\ngspc\to\ngrep$ (recursive code)
	\item $\ngspc\to\ngrep\to\ngbc\to\ngrep\to\ngspc$ (non-recursive code)
\end{itemize}

In this work the basic code node \ngbc is a rate-1/2 REP node \ngrep.
Each node update in the schedule computes LLR messages $\llrmsg^{\beta\to\gamma}_n$ from node $\beta$ to node $\gamma$ using extrinsic input messages $\llrmsg^{\alpha\to\beta}_n$ for $n=1,\ldots,\NC$. The following update rules are used
\begin{align}
	\ell^{\beta \to \gamma}_n = \begin{dcases}
		\sum_{\alpha \in \mathcal{N}(\beta)\setminus \gamma} \ell^{\alpha \to \beta}_n & \text{if $\beta$ is a REP node \ngrep}\\
		\myboxplus_{\alpha \in \mathcal{N}(\beta)\setminus \gamma} \ell^{\alpha \to \beta}_n & \text{if $\beta$ is an SPC node \ngspc}
	\end{dcases}\label{eq:rep_spc_updates}
\end{align}
where $\mathcal{N}(\beta)$ denotes the set of all nodes connected to $\beta$.
The boxplus operation is defined as
\begin{align}
	\ell_1 \boxplus \ell_2\mydef \operatorname{sgn}(\ell_1)\operatorname{sgn}(\ell_2)(\min(|\ell_1|, |\ell_2|){-}\mu(|\ell_1|,|\ell_2|))\label{eq:boxplus}&\\
	\mu(|\ell_1|,|\ell_2|)\mydef\log\frac{1+\exp(-||\ell_1|-|\ell_2||)}{1+\exp(-(|\ell_1|+|\ell_2|))}
\end{align}
where $\mu$ is a non-negative correction function.

\section{Properties of Recursive/Non-Recursive BMST}\label{sec:properties_rec_nrec}
\subsection{Error Propagation}\label{sec:error_propagation}
For rBMST, the feedback path in \cref{fig:encoders}b causes any bit error in $\xvec_t$ to propagate indefinitely through the recursive structure.
In contrast, the nBMST encoder shown in \cref{fig:encoders}c limits the effect of a bit error in $\xvec_t$ to $m$ subsequent blocks.
This behavior indicates that the recursive structure is more prone to error propagation.

From the decoder perspective, we need to consider processing of LLR messages.
In \cref{fig:ng_rec_nrec}, an edge case is shown where the decoding window covers the layers $t$ to $t+2$.
Consider one of the $N$ LLRs originating from layer $t-2$ to be erroneous denoted by \drawllrerror.
This LLR will not be corrected, as layer $t-2$ is outside the window.

In the rBMST graph (cf. \cref{fig:ng_rec_nrec}a) at $t$, the SPC node \ngspc propagates the erroneous LLR \drawllrerror if the remaining input LLRs \drawllrcorrect are correct.
Thus, the basic code node \ngbc and REP node \ngrep receive erroneous LLRs \drawllrerror, which may propagate into later layers.

In the non-recursive graph (cf. \cref{fig:ng_rec_nrec}b), an erroneous LLR \drawllrerror from layer $t-2$ can be corrected by the REP node \ngrep update rule in \cref{eq:rep_spc_updates} before being propagated to the basic code node \ngbc or SPC nodes \ngspc.

Thus, in the non-recursive structure, an erroneous LLR from a previous layer can be corrected, whereas in the recursive structure it propagates to multiple connected nodes.

\newcommand{\spclabel}{\mathtt{s}}
\newcommand{\replabel}{\mathtt{r}}
\newcommand{\bclabel}{\mathtt{c}}
\subsection{Error Correction Capability}\label{sec:error_correction}
Although less prone to error propagation, the non-recursive structure has significant limitations in terms of error correction capability.
Let $\spclabel$, $\replabel$, and $\bclabel$ denote SPC \ngspc, REP \ngrep, and basic code \ngbc nodes, respectively.
The channel messages $\ell^{\chlabel\to\spclabel}$ cannot improve during the iterations.
Hence, the LLR magnitudes $|\ell^{\spclabel\to\replabel}|$ are bounded by the channel LLR magnitudes due to the boxplus operation in \cref{eq:boxplus}.
In turn, the LLR magnitudes $|\ell^{\replabel\to\bclabel}|$ are bounded by the sum of the corresponding channel LLR magnitudes due to the REP update rule in \cref{eq:rep_spc_updates}:
\begin{align}
	\max(|\ell^{\replabel\to\bclabel}|)=\sum_{\spclabel\in\mathcal{N}(\replabel)\setminus\bclabel}|\ell^{\chlabel\to\spclabel}|\label{eq:ch_constraint}
\end{align}

In the case of a rate-$\nicefrac{1}{2}$ REP basic code, two LLR messages $\ell^{\replabel_1\to\bclabel}$ and $\ell^{\replabel_2\to\bclabel}$ are involved in the information bit decision:
\begin{align}
\hat{u}=\begin{dcases}
	0 & \text{for }\ell^{\replabel_1\to\bclabel}+\ell^{\replabel_2\to\bclabel} > 0\\
	1 & \text{otherwise}
\end{dcases}
\end{align}

An erroneous LLR with magnitude $|\ell^{\replabel_2\to\bclabel}|{>}\max(|\ell^{\replabel_1\to\bclabel}|)$ cannot be corrected by a correct LLR $\ell^{\replabel_1\to\bclabel}$.

The recursive structure, on the other hand, does not exhibit the limitation in \cref{eq:ch_constraint}.
The LLRs exchanged between REP and SPC nodes can grow across iterations without being bounded by $m+1$ channel messages.

\section{Mixed BMST via Parallel Concatenation}\label{sec:mixed_bmst_rcn}
\begin{figure}
	\centering	
	\includestandalone[mode=buildmissing]{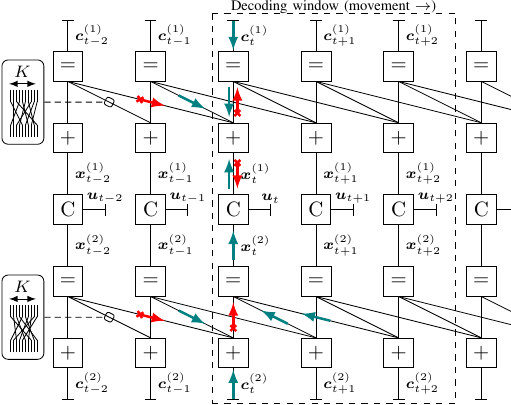}
	\caption{Error propagation in a mixed BMST graph. The non-recursive branch~2 is less prone to propagating errors into the recursive branch 1.}
	
	\label{fig:graph_rcn}
\end{figure}
\begin{figure}[t]
	\centering	
	\includestandalone[mode=buildmissing]{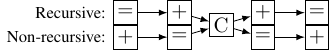}
	\caption{Schedule for updating a layer under mixed BMST decoding.}
	\label{fig:mixed_dec_schedule}
\end{figure}

The following proposes a structure that aims at combining the strengths of the recursive and non-recursive structures discussed in \cref{sec:properties_rec_nrec}.
The scheme is a parallel concatenation of the convolutional rate-1 codes.
After encoding $\uvec_t$ to $\xvec_t$ with the basic code, a branch operation $b$ divides $\xvec_t$ into $B$ segments $\xvec^{(b)}_t$.

In this letter, we restrict attention to the exemplary case $B=2$. Since a rate-$\nicefrac{1}{2}$ REP basic code is employed, the segments can be defined as $\xvec_{t}^{(1)}=\uvec_t$ and $\xvec_{t}^{(2)}=\uvec_t$.
The first branch is encoded using a recursive code, whereas the second branch is encoded using a non-recursive code:
\begin{align}
	\cvec_t=\begin{bmatrix}
		\cvec_t^{(1)}\\
		\cvec_t^{(2)}		
	\end{bmatrix} = 
	\begin{bmatrix}
		\coder_{m_1}(\xvec_t^{(1)})\\
		\coden_{m_2}(\xvec_t^{(2)})
	\end{bmatrix}
	\label{eq:rcn_codeword}
\end{align}

\Cref{fig:graph_rcn} shows the corresponding normal graph.
\Cref{fig:mixed_dec_schedule} describes a schedule for a layer update, where the recursive schedule can run concurrently with the non-recursive schedule.
This structure may improve implementation efficiency:
Updating a single LLR requires either a REP or an SPC computation module, as defined in \cref{eq:rep_spc_updates}.
In the mixed structure, each layer comprises two branches with $\KC$ parallel edges each.
A fully parallel implementation therefore requires $\KC$ REP modules and $\KC$ SPC modules, which can operate concurrently.
By contrast, the recursive structure in \cref{fig:ng_rec_nrec}a has a single branch per layer with $\NC$ parallel edges.
Consequently, a fully parallel layer update requires $\NC$ REP modules and $\NC$ SPC modules, which are executed sequentially.
Assuming identical delay for all modules, this implies that the mixed structure reduces the required implementation area by  $\nicefrac{K}{N}=\nicefrac{1}{2}$ while achieving the same throughput.
An additional advantage is that LLRs in the non-recursive structure may be represented with fewer bits under quantization, since their dynamic range is limited by the channel messages (cf. \cref{sec:error_correction}).
Moreover, permutations of size $\KC$ are less costly than permutations of size $\NC$.

\begin{figure}
	\centering
	\centering\includestandalone[mode=buildmissing]{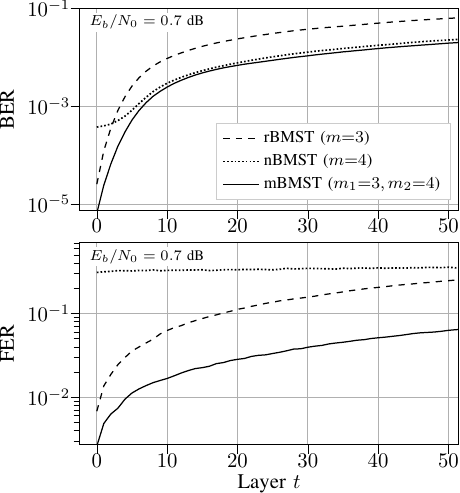}
	\caption{BER/FER per layer (info block length $\KC{=}1000$, code rate $\RC=\nicefrac{1}{2}$, coupling length $L{=}500$, window size $W{=}11$, termination length $T{=}W{-}1$, iteration count $\IW{=}10$).}
	\label{fig:ber_per_layer}
\end{figure}
\subsection{Numerical Analysis}
\subsubsection{Error Propagation Behavior}
\Cref{fig:ber_per_layer} compares the bit error rate (BER) and frame error rate (FER) for recursive (rBMST), non-recursive (nBMST), and mixed (mBMST) configurations.
A frame error is counted for layer $t$ if at least one bit error occurs in the estimate of the information block $\uvec_t$ after the sliding window has completed all iterations.
For all configurations, the BER and FER are lowest at the beginning, since the initial symbols stored in the encoder memory are known.
As the layer index $t$ increases, both metrics degrade due to error propagation from previous layers.

For nBMST, the BER is lower than for rBMST, while the FER is higher.
This indicates that frame errors typically contain only a small number of bit errors.
A contributing factor is that sequences of low-magnitude channel LLRs can prevent error correction, as discussed in \cref{sec:error_correction}.
In contrast, rBMST exhibits a lower FER but a higher BER, implying that frame errors tend to contain many bit errors.
This behavior can be attributed to recursive error propagation, where a single bit error entering the next layer can trigger additional errors, as discussed in \cref{sec:error_propagation}.
The proposed mBMST configuration balances these effects and achieves the lowest BER and FER across all layers.

\subsubsection{Iterative Behavior}
This section studies how the quality of LLR messages evolves across decoding iterations.
From the perspective of a layer $t$, the sliding window gradually moves over it.
At each window position, the layer undergoes two updates in each of the $\IW$ iterations.
Exceptions occur when the layer $t$ is the leftmost or rightmost layer in the sliding window, where only a single layer update is applied per iteration.
For a window size $W$, a layer therefore experiences $2\IW(W-1)$ updates.

\begin{figure}
	\centering
	\includestandalone[mode=buildmissing]{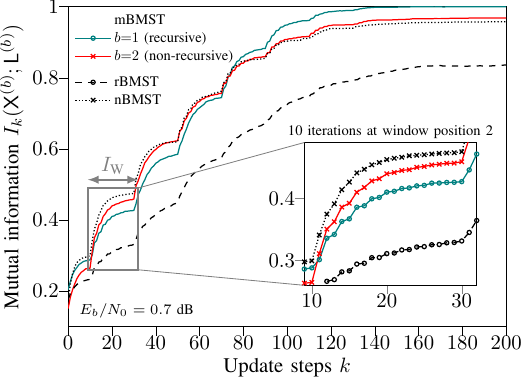}
	\caption{Evolution of mutual information for layer $t=25$ from \cref{fig:ber_per_layer}.}
	\label{fig:mi_evolution_for_layer}
\end{figure}

Let $\myvec{\ell}^{(b)}_t$ denote the LLR message vector generated by branch~$b$ which is input to the basic code node \ngbc.
The LLRs in $\myvec{\ell}^{(b)}_t$ and the corresponding code bits in $\myvec{x}^{(b)}_t$ can be regarded as realizations of random variables $\LL^{(b)}$ and $\XX^{(b)}$, respectively, whose statistics are estimated empirically from simulation data.
\Cref{fig:mi_evolution_for_layer} shows the mutual information $I_k(\XX^{(b)},\LL^{(b)})$ for the mBMST, rBMST, and nBMST decoders from \cref{fig:ber_per_layer}, evaluated for layer $t=25$.
The update steps are indexed by $k=1,\dots,2\IW(W-1)$. LLRs $\myvec{\ell}^{(b)}_t$ are called \emph{informative} if $I_k(\XX^{(b)},\LL^{(b)})>0$.

We first analyze the mutual information evolution for the two branches of the mBMST configuration.
As specified in \cref{eq:rcn_codeword}, the recursive component corresponds to branch~1 and the non-recursive component to branch~2.
For steps 1 to 10, layer $25$ is the right-most layer in the sliding window, similarly as the layer $t+2$ in \cref{fig:graph_rcn}.
In this range, the mutual information of branch~2 is lower than that of branch~1.
This is because only the first of the $m+1$ connected SPC nodes to the REP node in branch~2 provides informative LLRs.
These LLRs are bounded by the channel LLRs entering the SPC node.
In contrast, the SPC node in branch~1 is connected to $m+1$ REP nodes, each contributing informative LLRs.
Once the window advances by one position, the REP node in branch~2 receives two informative LLRs.
As a result, $I_k(\XX^{(2)},\LL^{(2)})$ surpasses $I_k(\XX^{(1)},\LL^{(1)})$ at step $10$.
After step $80$, a second crossing point occurs where $I_k(\XX^{(1)},\LL^{(1)})$ becomes dominant, since the LLR magnitudes from branch~2 are bounded as discussed in \cref{sec:error_correction}.
Ultimately, only $I_k(\XX^{(1)},\LL^{(1)})$ converges to $1.0$, whereas $I_k(\XX^{(2)},\LL^{(2)})$ saturates at approximately $0.95$ for the chosen $\EbNo$.

Looking at the conventional configurations, the rBMST decoder does not converge, while the nBMST decoder exhibits high mutual information initially but also saturates, similar to branch~2 in the mBMST configuration.

\section{Parallel and Serial Concatenation}

\subsection{Generalization}
In addition to parallel concatenation with branches $b \in \{1,\ldots,B\}$, serial concatenation with stages $s \in \{1,\ldots,S_b\}$ can be applied within each branch.
The output of stage $s+1$ in branch $b$ is given by
\begin{align}
	\xvec_t^{(b,s+1)} \mydef \codegeneral_{b,s}(\xvec^{(b,s)}_t),
\end{align}
where the initialization at stage~1 is $\xvec_t^{(b,0)} = \xvec^{(b)}_t$, and the output of stage~$S_b$ is denoted by $\cvec_t^{(b)} = \xvec_t^{(b,S_b)}$.
Each component code $\codegeneral_{b,s}$ is characterized by a memory length $m_{b,s}$ and operates either in recursive or non-recursive mode:
\begin{align}
	\codegeneral_{b,s}(\xvec_t^{(b,s)})
	= \begin{dcases}
		\coder_{m_{b,s}}(\xvec_t^{(b,s)}) & \text{recursive},\\
		\coden_{m_{b,s}}(\xvec_t^{(b,s)}) & \text{non-recursive}.
	\end{dcases}
	\label{eq:components_codes_b_s}
\end{align}
The overall codeword is then obtained as
\begin{align}
	\cvec_t =
	\begin{bmatrix}
		\cvec_t^{(1)}\\
		\vdots\\
		\cvec_t^{(B)}
	\end{bmatrix}
	=
	\begin{bmatrix}
		\codegeneral_{1,S_1}(\ldots(\codegeneral_{1,1}(\xvec^{(1)}_t)))\\
		\vdots\\
		\codegeneral_{B,S_B}(\ldots(\codegeneral_{B,1}(\xvec^{(B)}_t)))
	\end{bmatrix}.
	\label{eq:mbmstgeneralcodewords}
\end{align}

\begin{figure}[t]
	\centering
	\includestandalone[mode=buildmissing]{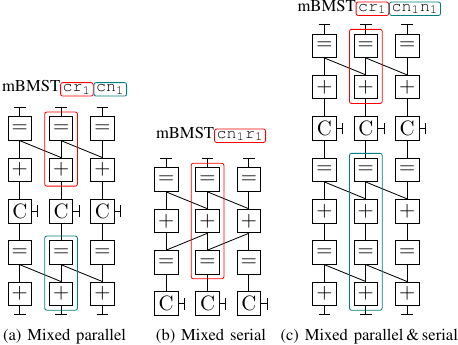}
	\caption{Mixed configurations with recursive and non-recursive components.}
	\label{fig:mbmstconfigurations}
\end{figure}

\Cref{fig:mbmstconfigurations} depicts the graphs corresponding to several possible configurations defined by \cref{eq:mbmstgeneralcodewords} with $B\leq 2$.
The naming scheme, such as \mBMST{cr1cn1n1}, provides a compact representation of the code structure.
The symbol \mBMST{c} denotes the beginning of a branch starting from a basic \underline{c}ode.
A serial concatenation of \underline{r}ecursive and \underline{n}on-recursive components within a branch is specified by a sequence of \mBMST{r} and \mBMST{n} symbols.
An additional \mBMST{c} indicates the presence of a second branch corresponding to parallel concatenation.
Optional subscripts specify the corresponding memories $m_{b,s}$.
Existing schemes such as BMST\cite{ma2015block}, rBMST~\cite{zhao2018recursive}, DrBMST~\cite{zhaodoublyrecursive2020}, and BiBMST~\cite{li2024spatially} are denoted by \mBMST{cn}, \mBMST{cr}, \mBMST{crr} and \mBMST{crn}, respectively.
New configurations use parallel concatenation (e.g. \mBMST{crcn}), serial concatenation (e.g. \mBMST{cnr}) or parallel \& serial concatenation (e.g. \mBMST{crcnn}). 
\begin{figure*}[t]
	\centering
	\includestandalone[mode=buildmissing]{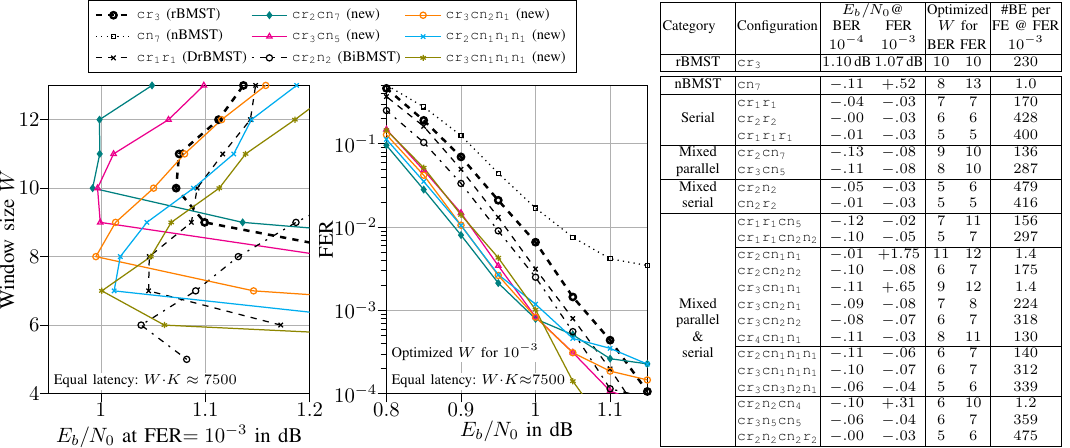}
	\caption{BER/FER of mixed BMST configurations. The decoding latency $W\cdot K$ is kept constant. FER is evaluated with frame length of $1000$.}
	\label{fig:ber_mixed_vs_conventional}
\end{figure*}
\subsection{Performance Analysis}
For fair comparisons, the BER and FER performance of different configurations is compared at constant decoding latency $W\cdot\KC$.
The ratio $\nicefrac{W}{\KC}$ can be optimized for each configuration.
In addition the coupling length $L$ is chosen such that the rate satisfies $R=\RC (\nicefrac{L}{L+T})=0.49$ with termination length $T=W-1$.

In \cref{fig:ber_mixed_vs_conventional} we provide and overview of the performance with many configurations.
A strong dependence on the window size $W$ can be noticed.
A serial concatenation tends to favor smaller $W$ whereas the parallel concatenation benefits from larger $W$.
Furthermore, the table in \cref{fig:ber_mixed_vs_conventional} shows, that optimizing for BER performance leads to smaller $W$ than optimizing for FER performance.

The baseline configuration, \mBMST{cr3} (rBMST), achieves an $\text{FER}$ of $10^{-3}$ at $\EbNo=1.07\dB$ and a $\text{BER}$ of $10^{-4}$ at $\EbNo=1.10\dB$.
The \mBMST{cn7} (nBMST) decoder can achieve better BER performance, but requires substantially more memory and still exhibits a high FER error floor. 
At FER of $10^{-3}$, each erroneous frame contains a single bit error (see the last column of the table), resulting from the limited error correction capability discussed in \cref{sec:error_correction}.

A significantly improved memory-performance trade-off is achieved using mixed configurations.
For example, \mBMST{cr2cn7} requires only a memory of 2 in the first branch and improves the BER and FER performance by $0.13\dB$ and $0.08\dB$, respectively.
The memory requirements can be further reduced through serial concatenation in the non-recursive branch.
For example, compared to \mBMST{cr2cn7}, \mBMST{cr2cn1n1n1} reduces non-recursive memory from 7 to 3, with comparable performance.
Increasing memory in the recursive branch, as in \mBMST{cr3cn1n1n1}, can improve the error floor performance.
Other promising candidates use a decreasing memory in the non-recursive branch, e.g., \mBMST{cr3cn2n1}.
However, the memory assignment must be well balanced with the recursive branch.
For example, \mBMST{cr3cn3n2n1} performs significantly worse than \mBMST{cr2cn1n1n1}.

Configurations available in the literature, such as \mBMST{cr1r1} (DrBMST) \cite{zhaodoublyrecursive2020} and \mBMST{cr2n2} (BiBMST) \cite{li2024spatially}, achieve FER gains of $0.03\,\mathrm{dB}$ over rBMST. 
In comparison, the proposed configurations achieve an FER improvement of $0.08\,\mathrm{dB}$, representing a noticeable gain over the state of the art.

\section{Conclusions}
A new class of BMST variants, termed mixed BMST (mBMST) codes, combines recursive and non-recursive component codes through parallel and serial concatenation.
Parallel mixing yields performance improvements of up to 0.13\,dB compared to rBMST codes.
Serial concatenation enables a reduction in encoding memory complexity while maintaining comparable performance.
Open issues include the impact of the choice of the basic code, such as its rate and strength, as well as the behavior of systems with more than two branches.
Also exploiting the mixed schedule in hardware implementations remains an important direction for future work.

\bibliographystyle{IEEEtran} %
\bibliography{reference}

\end{document}

%% file: preamble_tikz.tex
\input{commands.tex}
\usepackage{amsmath} %
\usepackage{mathtools} %
\usepackage{tikz}
\usepackage{pgfplots}
\usepackage{currfile}
\usepackage{ifthen}
\usepackage{xspace}
\usepackage{makecell}

\usepackage{xparse}
\usepackage{expl3}
\usepackage{multirow}
\usetikzlibrary{positioning, calc,arrows.meta, fit}

\tikzset{
	errorarrow/.style={red,
		{Rays[width'=0pt 1, width=4pt,length=4pt]}-latex, very thick
	}
}

\tikzset{
	correctarrow/.style={teal,-latex, very thick}
}

\newcommand{\shiftreg}[1]{
	\def\sym_reg_inp{\ifnum#1=0 x\else c\fi}
	\def\boxdist{0.4}		
	\def\Dboxfactor{1.3}
	\def\memory{2}
	\foreach \m in {1,...,\memory}{
		\pgfmathsetmacro{\sign}{#1==0 ? 1 : -1}
		\node(D\m)[draw=black] at (\sign*\m*\Dboxfactor,0){$D$};
		\node(perm\m)[draw=black, above=\boxdist of D\m, inner sep=2pt]{$\myvec{\mathrm{\Pi}}_\m$};
	}
	\coordinate (helper) at ($(perm1.north)!0.5!(perm\memory.north)$);
	\node[draw=black, inner sep=3pt, minimum width=2cm,above=\boxdist of helper] (sum) {$\oplus$};
	\node(x)[left=0.8 of sum]{};
	\node(c)[right=0.8 of sum]{};
	\draw [-latex](x)--(sum.west)node[above, pos=0.5]{$\myvec{x}_t$};
	\draw [-latex](sum.east)--(c)node[above, pos=0.5]{$\myvec{c}_t$};

	\foreach \m in {1,...,\memory}{
		\draw[-latex] (D\m)--(perm\m)node[right, pos=0.5]{$\myvec{\sym_reg_inp}_{t-\m}$};
		\draw[-latex] (perm\m)--(perm\m|-sum.south);
	}
	\foreach \m in {2,...,\memory}{					
		\pgfmathtruncatemacro{\mbefore}{\m-1}
		\draw[-latex] (D\mbefore)--(D\m);
	}
	\ifthenelse{\equal{#1}{0}}
	{\coordinate (helper) at ($(sum.west)!0.5!(x)$);}
	{\coordinate (helper) at ($(sum.east)!0.5!(c)$);}		
	\draw[-latex] (helper)|-(D1);
}

%% file: commands.tex
\newcommand\mylabel[1]{\text{#1}}
\newcommand{\CBasic}{\mathrm{C}}

\newcommand{\RC}{R_\mathrm{C}}
\newcommand{\IW}{I_\mylabel{W}}

\newcommand{\Pmat}{\mathbf{\Pi}} %

\newcommand{\KC}{K}
\newcommand{\NC}{N}

\newcommand{\mydef}{\triangleq}

\input{ng_sym_helpers_old.tex}

\input{ng_sym_helpers.tex}
\DeclareMathOperator{\sign}{sgn}

\newcommand{\myboxplus}{\mathop{\vphantom{\bigoplus}\mathchoice
		{\vcenter{\hbox{\resizebox{\widthof{$\displaystyle\bigoplus$}}{!}{$\boxplus$}}}}
		{\vcenter{\hbox{\resizebox{\widthof{$\bigoplus$}}{!}{$\boxplus$}}}}
		{\vcenter{\hbox{\resizebox{\widthof{$\scriptstyle\oplus$}}{!}{$\boxplus$}}}}
		{\vcenter{\hbox{\resizebox{\widthof{$\scriptscriptstyle\oplus$}}{!}{$\boxplus$}}}}
	}\displaylimits
}

\newcommand\myrv[1]{\mathsf{#1}}
\newcommand{\LL}{\myrv{L}}
\newcommand{\XX}{\myrv{X}}

\newcommand{\myvec}[1]{\boldsymbol{#1}}
\newcommand{\cvec}{\myvec{c}}
\newcommand{\uvec}{\myvec{u}}
\newcommand{\xvec}{\myvec{x}}
\newcommand{\nvec}{\myvec{n}}
\newcommand{\yvec}{\myvec{y}}
\newcommand{\lchvec}{\myvec{\ell}^{\mylabel{ch}}}
\newcommand{\chlabel}{\mylabel{ch}}

\newcommand{\No}{N_0}
\newcommand{\EbNo}{E_b/N_0}

\newcommand\llrmsg{\ell}

\newcommand\labelr{\mylabel{r}}
\newcommand\labeln{\mylabel{n}}

\newcommand{\codegeneral}{\CBasic}
\newcommand{\coder}{\CBasic^\labelr}
\newcommand{\coden}{\CBasic^\labeln}

\makeatletter
\newcommand{\mbmst}[1]{%
	\mbmst@list#1-\@nil
}
\def\mbmst@list-#1-#2\@nil{%
	\mbmst@item#1\@nil
	\if\relax\detokenize{#2}\relax
	\else
	-\mbmst@list-#2\@nil
	\fi
}
\def\mbmst@item#1#2\@nil{%
	\ifcat a#1%
	#1\mbmst@item#2\@nil
	\else
	\textsubscript{#1#2}%
	\fi
}
\makeatother

\newcommand{\drawllrerror}{\tikz[baseline={($(X.center)+(0pt,-2.7pt)$)}]\draw[errorarrow] (0,0)--(0.5,0);\xspace}
\newcommand{\drawllrcorrect}{\tikz[baseline={($(X.center)+(0pt,-2.7pt)$)}]\draw[correctarrow] (0,0)--(0.5,0);\xspace}

\ExplSyntaxOn

\NewDocumentCommand{\mBMST}{m}
{
	\mbox{%
		\texttt{%
			\tl_set:Nn \l_tmpa_tl { #1 }
			\regex_replace_all:nnN
			{ ([0-9]+) }
			{ \c{textsubscript}\cB{\1\cE} }
			\l_tmpa_tl
			\tl_use:N \l_tmpa_tl
		}%
	}
}

\ExplSyntaxOff

\newcommand{\dB}{\,\mathrm{dB}}

%% file: ng_sym_helpers_old.tex
\newcommand{\ngrep}{\tikz[baseline={($(X.center)+(0pt,-2.7pt)$)}]\node[draw, inner sep=1pt, minimum size=0.35cm, fill=white, anchor=center, align=center,label={[yshift=-0.8pt, inner sep=0pt]center:$=$}] (X) {};\xspace}
\newcommand{\ngspc}{\tikz[baseline={($(X.center)+(0pt,-2.7pt)$)}]\node[draw, inner sep=1pt, minimum size=0.35cm, fill=white, anchor=center, align=center] (X) {$+$};\xspace}
\newcommand{\ngbc}{\tikz[baseline={($(X.center)+(0pt,-2.7pt)$)}]\node[draw, inner sep=1pt, minimum size=0.35cm, fill=white, anchor=center, align=center] (X) {$\mathrm{C}$};\xspace}